\documentclass[sigconf]{acmart}

\usepackage{xcolor}

\usepackage{acmart-taps}
\usepackage[]{algorithm2e}
\usepackage{mathrsfs}
\usepackage{enumitem}
\usepackage{mathtools}
\usepackage{amsthm}

\usepackage{pgfplots, pgfplotstable, booktabs, colortbl, siunitx, array}
\usepackage{subcaption}
\usepackage{xspace}
\usepackage{dsfont}  

\newcolumntype{R}[1]{>{\raggedleft\arraybackslash }b{#1}}

\newcommand{\betad}{\textsc{Beta\xspace}}
\newcommand{\scaledbetad}{\textsc{ScaledBeta\xspace}}

\newcommand{\bidorders}{\textsc{BidOrders\xspace}}
\newcommand{\askdorders}{\textsc{AskOrders\xspace}}

\newcommand{\bid}{\texttt{bid\xspace}}
\newcommand{\ask}{\texttt{ask\xspace}}
\newcommand{\inv}{\texttt{inv\xspace}}
\newcommand{\minquote}{\texttt{min\_quote\xspace}}
\newcommand{\maxinv}{\texttt{max\_inv\xspace}}
\newcommand{\clamp}{\texttt{clamp\xspace}}

\newcommand{\defomega}{\omega_0}
\newcommand{\defkappa}{\kappa_0}
\newcommand{\maxkappa}{\kappa_{\max}}
\newcommand{\invfrac}{\texttt{frac\_inv\xspace}}
\newcommand{\nlevels}{\texttt{n}\hbox{$\_$}\texttt{levels}\xspace}

\newcommand{\totalvolume}{\texttt{total\_volume\xspace}}
\newcommand{\ticker}[1]{\texttt{#1}}

\newcommand{\abid}{\ensuremath{\alpha^\texttt{\scalebox{0.8}{bid\xspace}}}}
\newcommand{\bbid}{\ensuremath{\beta^\texttt{\scalebox{0.8}{\kern 0.06667em bid\xspace}}}}
\newcommand{\aask}{\ensuremath{\alpha^\texttt{\scalebox{0.8}{ask\xspace}}}}
\newcommand{\bask}{\ensuremath{\beta^\texttt{\scalebox{0.8}{\kern 0.06667em ask\xspace}}}}
		
\AtBeginDocument{%
  \providecommand\BibTeX{{%
    \normalfont B\kern-0.5em{\scshape i\kern-0.25em b}\kern-0.8em\TeX}}}

\copyrightyear{2022}
\acmYear{2022}
\setcopyright{acmlicensed}
\acmConference[ICAIF '22]{3rd ACM International Conference on AI in Finance}{November 2--4, 2022}{New York, NY, USA}
\acmBooktitle{3rd ACM International Conference on AI in Finance (ICAIF '22), November 2--4, 2022, New York, NY, USA}
\acmPrice{15.00}
\acmDOI{10.1145/3533271.3561745}
\acmISBN{978-1-4503-9376-8/22/10}

\begin{document}

\title{Market Making with Scaled Beta Policies}

\author{Joseph Jerome}
\affiliation{%
	\institution{Department of Computer Science}
  \streetaddress{University of Liverpool}
  \city{Liverpool}
  \country{UK}
}
\email{j.jerome@liverpool.ac.uk}

\author{Gregory Palmer}
\affiliation{%
	\institution{L3S Research Center}
  \streetaddress{Leibniz University Hannover}
  \city{Hannover}
  \country{Germany}
}
\email{gpalmer@l3s.de}

\author{Rahul Savani}
\orcid{0000-0003-1262-7831}
\affiliation{%
	\institution{Department of Computer Science}
  \streetaddress{University of Liverpool}
  \city{Liverpool}
  \country{UK}
}
\email{rahul.savani@liverpool.ac.uk}

\begin{abstract}
This paper introduces a new representation for the actions of a market
maker in an order-driven market.
This representation uses scaled beta distributions, and generalises three
approaches taken in the artificial intelligence for market making literature:
single price-level selection, ladder strategies, and ``market making at the
touch''.
Ladder strategies place uniform volume across an interval of 
contiguous prices.
Scaled beta distribution based policies generalise these, allowing volume to be skewed across the price interval.
We demonstrate that this flexibility is useful for inventory management,
one of the key challenges faced by a market maker.
We conduct three main experiments: first, we compare our more flexible
beta-based actions with the special case of ladder strategies; then, we
investigate the performance of simple fixed distributions; and finally, we
devise and evaluate a simple and intuitive dynamic control policy that adjusts
actions in a continuous manner depending on the signed inventory that the market
maker has acquired.
All empirical evaluations use a high-fidelity limit order book simulator based on 
historical data with 50 levels on each side.
\end{abstract}





\maketitle

\section{Introduction}


This paper considers the problem of a market maker acting in an order-driven market.
In such markets, matched orders result in trades and unmatched orders are stored
in a \emph{limit order book}, which is split into two parts, a collection of 
buy orders called \emph{bids}, and a collection of sell orders called \emph{asks}.
A market maker provides liquidity by continuously having both bids and asks
in the book, thereby allowing others to trade in either direction whenever needed.
The difference between the market maker's best ask and best bid is called their 
\emph{quoted spread}, which could be wider and use different prices than the spread of 
the whole limit order book market.
The goal of a market maker is to repeatedly earn this spread
by transacting in both directions.
The challenge for the market maker is to mitigate the inventory risk
that comes from trading with better-informed traders.
That is, market makers expose themselves to
\emph{adverse selection}, a phenomenon where the market maker's counterparties exploit a technological or
informational advantage when transacting with them. In particular, this causes the market maker
to amass a (positive or negative) inventory, before 
an adverse price move causes the market maker to incur a loss on this inventory
(for example, where the market maker has net sold to the market just before a 
significant price rise).

Market making has become increasingly automated and the frequency of trading and
corresponding data requirements has grown significantly
~\cite{Hasbrouck2013,Leaver2016,Savani2012,Haynes2015}.
This paper investigates a novel but natural way to represent the actions of 
an automated market maker. 
Our approach uses scaled beta distributions as a flexible and succinct way
to define the volume profiles of bids and asks that a market maker places.
We demonstrate the utility of this representation by backtesting and analysing
market making agents that use this representation within high-fidelity simulations
of the limit order book (using LOBSTER\footnote{\url{https://lobsterdata.com/}}
data with 50 levels of limit orders on each side of the book).

\begin{figure}[h]
\centering
\includegraphics[width=\linewidth]{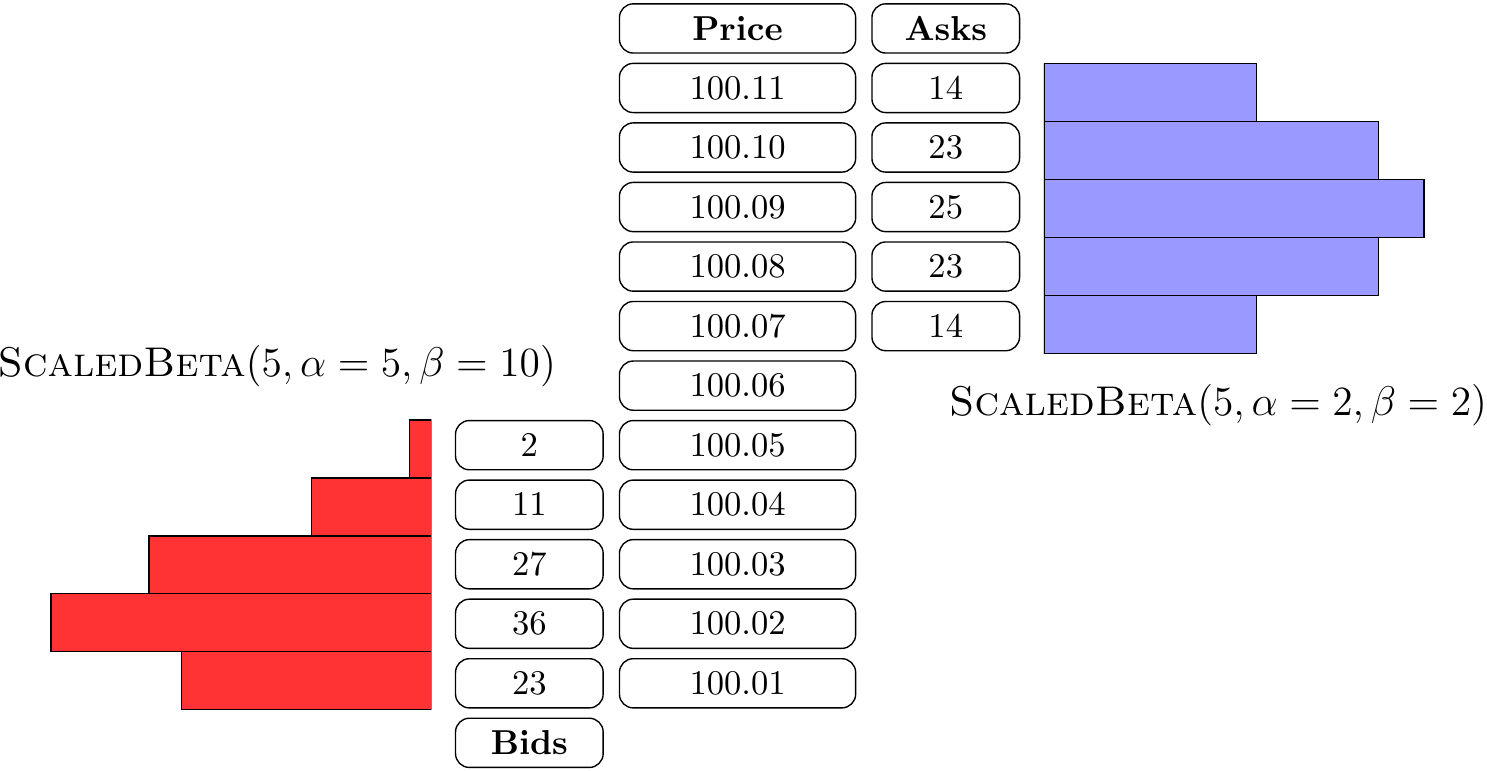}
\caption{An example of a limit order book, with an illustration of order 
distributions described by beta distributions. 
}
\label{fig:lob}
\end{figure}

Figure~\ref{fig:lob} gives an illustration of how we use beta distributions to
describe volume profiles for orders across price levels. For the sake of
illustration, assume that all shown orders come from the same market marker. The
bid and ask volumes in this example are derived from $\scaledbetad(5,5,10)$ and
$\scaledbetad(5,2,2)$, where the first argument, 5, is the number of levels to
quote at, and the second and third parameters are the shape parameters of a standard beta
distribution ($\alpha$ and $\beta$ respectively; see Section~\ref{subsec:beta}).

\subsection{Related work}\label{sec:literature}

Market making has been investigated within the economics, finance, and artificial
intelligence (AI) literatures. The classic approach taken in the mathematical finance literature has been to treat market
making as a problem of \emph{stochastic optimal control}, where models for order
arrivals and executions are developed and then control problems are designed and solved
for them~\cite{ho1981optimal,grossman1988liquidity,avellaneda2008high,Guilbaud2011,CJR14,abergel2016limit,gueant2013dealing,Cartea2015,JSSHS22}.
Because of the emphasis on analytically proving results about optimal or approximately
optimal control policies, the action space of the market maker is typically quite restrictive.
This often happens to the point where the market
maker only controls a single order on each side of the market and therefore a
single spread. The main novelty of this paper is a flexible parametric
representation of order profiles, which permits order placement across many prices.

We give a brief summary of the AI for
market making (AI4MM) literature. Since the main focus of this paper is the
novel policy, we predominantly contrast this with the policies used in other
parts of the AI4MM literature, where, broadly speaking, the action spaces can be
divided into three categories: that of choosing discrete half spreads
at which to place buy and sell orders from a finite set (whilst possibly also
managing the amount of volume placed at both levels); that of choosing a range
for a ``ladder strategy''; and that of ``market-making at-the-touch'' where the
actions consist in choosing either to place an order or not at each of the best
bid and best ask. We discuss these three approaches in turn.

\subsubsection{Single price-level policy}\label{sec:single-price-level}
A natural two-dimensional action space is for the agent to select two
\emph{half-spreads}, i.e. a bid and ask offset from the midprice. 
Typically, in the literature, with this setup all orders are assumed to be 
of constant volume.\footnote{Alternatively, one could select a volume at each of
these levels, making the action space of the control problem four-dimensional.}
The agent then adapts these half-spreads at each time step according to the
state of the market and the agent's inventory. 
This approach of choosing half spreads is the main one taken in the financial
stochastic control literature on market making~(e.g., see Avellaneda and
Stoikov~\cite{avellaneda2008high}, Gueant et al.~\cite{gueant2013dealing} or
Cartea, Jaimungal and co-authors~\cite{cartea2015algorithmic, CJR14,
cartea2017algorithmic}). 

While the financial stochastic control literature tends to use continuous models
with corresponding continuous half-spreads, by and large, the AI4MM 
literature restricts to the case where market makers have discrete action
spaces. In particular, the problem is that of choosing the \textit{number of
ticks} away from the touch (the best bid and ask prices) at which to quote the
bid and the ask. It is worth noting that due to the action space being the
product space of bid and ask actions, this can get very large unless the agent
is restricted to place actions very close to the best prices.

\paragraph{Reducing the action space by taking differences.}
The first application of reinforcement learning to market making
(\citet{Chan2001}) used such a policy. However, to get around the issue of the
large action space, they instead chose, from a small set of actions, how much they would increase or decrease
their quotes. They then used reinforcement
learning to optimise an agent's interaction with a mathematical model (similar
to that proposed by Glosten and Milgrom~\cite{glosten1985bid}) for the market
dynamics. Subsequently, Kim and Shelton~\cite{Kim02} fitted an input-output hidden
Markov model to order data from Nasdaq and use reinforcement learning to learn
how to act in the model. This was a significant improvement in terms of realism. 
Still, due to only observing a fraction of the order flow volume they still needed to
use a model of the market. They allowed their agent to increase,
decrease, or keep their bid, ask, and both associated volumes by
at most one (tick or unit of the asset). Whilst this seems rather
restrictive (as it requires many steps to make a dramatic 
change), the action space is already $(3^4 = 81)$-dimensional.

\paragraph{Choosing an action from a prespecified subset of available actions.}
More recently, Spooner et al.~\cite{spooner2018market} considered a much more
realistic market simulator, using 5 levels of order book data, along with
transactions. This was the first paper to train reinforcement learning
agents on the vast quantities of so-called \textit{Level II} data now available.
However, there is still a slight partial observability problem: when the order
book changes, and a transaction doesn't occur, it is not possible to know from
where in the queue this cancellation/deletion came from. In particular, when
interacting with the market, it is necessary to assume a distribution of such
cancellations. The authors of \cite{spooner2018market} chose a uniform
distribution. 

The action space of \citet{spooner2018market} is an octuple of pre-specified
half spread pairs, along with an action which clears the entirety of the agent's
inventory using a market order. It is worth noting that some of these actions
are skewed to favour filling on one side. This approach permits a basic form of
the inventory control discussed in Section \ref{sec:inventory-driven-policy},
whilst keeping the action space of manageable size. This paper was the first
to use such a finite pre-specified selection of actions and it has since been
utilised by \citet{XCH22performance} and \citet{Sadighian19}. Whilst
\citet{spooner2018market} used SARSA($\lambda$) and a state-space
discretisation, \cite{XCH22performance} used a variant of a deep Q networks and
Sadighian~\cite{Sadighian19} used proximal policy optimisation
\cite{SchulmanWDRK17}. At a similar time to \citet{spooner2018market},
\citet{LG18} considered a stochastic model driven by Poisson processes of the
form proposed by Cont et al~\cite{ContST10}, and allowed an agent to quote a
single bid and ask of fixed volume a number of ticks away from the best bid,
here chosen from the set $\mathscr{A} = \{1,2,3\}^2$. A similar policy is
adopted by Patel~\cite{patel2018optimizing}. However, to reduce the
dimensionality of the action space, only one half spread is chosen (on either 
side) and a limit order is placed at the best price on the other side of the book.

\paragraph{A continuous action space.}
Spooner and Savani~\cite{spooner2020robust} considered a robust version of the
Avellaneda and Stoikov model~\cite{avellaneda2008high} in which an adversarial
``market'' agent controls the drift of the financial market. This problem ends
up being equivalent to the problem considered by Nystr\"om et al.~\cite{NOAZ14}.
Here, the action space is given by four continuous parameters controlling the
mean and variance of the agent's bids and asks. It is learnt by approximating
the value function using cubic polynomials and then performing least squares
policy iteration~\cite{lagoudakis2003least}.

A final form of action space which falls into the \textit{single price-level}
category is that of choosing a continuous half spread on each side of the book and then 
quantising to submit orders that are on the price grid. This approach is taken by Ga\v
sperov and Kostanj\v car \cite{GasperovK21} in which they use neuroevolution to
train a policy given by a deep neural network. They use historical Bitcoin trade
data with only the first level of the order book.

It is worth noting that such policies based on half-spreads are a subclass of the
scaled beta policy that we introduce here (they can be recovered by letting the
variance decrease to zero). However, the authors feel that choosing half-spreads 
is not so sensible for the following reason: when actually implementing such a
strategy, to change the half spread on each side of the book, it is necessary to
actively cancel orders at each time step and place new orders at the new level.
Furthermore, even if the agent doesn't change their spread it is necessary to
cancel and place new orders to maintain a fixed spread (when the midprice
moves). This causes the agent to perpetually lose queue position on
price-time priority exchanges when they place new orders (at the back of
the queue). In particular, very few of their orders actually get filled. In
contrast, if an agent places orders according to a $\scaledbetad$ distribution,
their orders get updated in a smooth manner by constantly adding or
removing smaller orders to the book (see Figure~\ref{fig:bookdensity} for an
illustration).

\subsubsection{Ladder strategies.}
Another relevant strand of the existing literature was started by
\citet{chakraborty2011market}. In this paper, the authors introduced and studied the use of
\emph{ladder strategies}, which place a unit of volume
at all prices in two price intervals, one on each side of the book.
\cite{chakraborty2011market} theoretically proved the utility of these
strategies in mean-reverting markets with
Ornstein-Uhlenbeck price dynamics. 
Inspired by \cite{chakraborty2011market}, 
\citet{AbernethyK13} considered related order placement strategies where
limit orders for one unit of volume are placed at all price levels (right 
down to the lowest possible price and up to some predefined highest price)
outside of a window around the midprice that defines the market maker's spread.
\cite{AbernethyK13} presented an online learning scheme that mixes between
parametrisations of their ladder strategies\footnote{In doing
so, they allow trading fractional units, which is not realistic. However, their 
focus is primarily theoretical.} and in doing so provably guarantees to 
perform competitively with the single best parameter choice in hindsight.
They do an empirical evaluation with real data, but only using the price 
time series of trades, rather than actually modelling the limit order book
process as we do here. This requires one to make assumptions about fill-rates,
agent queue positions and many other aspects of the market's microstructure. That
being said, such a strategy deals with the problem mentioned at the end of
Section \ref{sec:single-price-level} of losing queue position, and we use it as
a benchmark throughout the paper. Both \cite{chakraborty2011market} and
\cite{AbernethyK13} are primarily
about the theoretical guarantees that their market making strategies provide,
whereas the focus in this paper is on exploring the utility of policies based
on our representation of order profiles in realistic high-fidelity simulations.
Moreover, the order profiles that our representations allow are much more 
flexible, generalising ladder strategies, which are recovered as a single 
parameter choice within our representation.

\subsubsection{Market-making at the touch}
Finally, it is worth discussing the problem of ``market-making at-the-touch''.
This problem is considered in Cartea et al.~\cite[Chapter
10.2.2]{cartea2015algorithmic} and Cartea et al.~\cite{CDJ2018} in a continous
time and space mathematical model of the market. In the reinforcement learning
literature, this approach is taken by \citet{ZBW20}. Here, the agent's actions
consist in choosing to have an order or not at both the touch of the bid and ask
sides of the book at every time step. By discretising the observation space
similarly to \cite{spooner2018market}, they achieve decent results with a
Q-learning agent. This extremely simple action space fares quite well, and
avoids the issues with the single price-level policy of
Section~\ref{sec:single-price-level}, provided the agent does not cancel and
replace orders too frequently. Again, this policy can effectively be recreated
using a beta policy, which we describe in detail in Section~\ref{sec:beta-generalisation}. 

\subsection{Our contributions} 

The key contributions of this paper are as follows:

\begin{itemize}[leftmargin=0.35cm]

\item We introduce a new parametric representation of market maker policies
	via scaled beta distributions. We show how
	these policies capture -- as special cases -- standard actions
	spaces from the literature, including single-level orders, ladder
	strategies, and market making at the touch. 
	The resulting continuous action space 
	is far more flexible than these special cases and allows the market maker
	to skew orders to address the problem of accumulated inventory 
        whilst maintaining their queue position.

\item We have developed a high-fidelity order book simulator, using LOBSTER data 
	which is combined with orders from our agents. We
	empirically evaluate our beta policies, showing first the benefit of
	non-uniform policies over ladder strategies.

\item We then explore inventory management. First we demonstrate the cost of
	using market orders to control inventory. Motivated by this we use our
	scaled beta policies to design an inventory-driven policy that
	automatically skews the distributions to favour driving the absolute value
	of inventory back towards zero. The policy controls inventory using only
	limit orders.
\end{itemize}
\noindent The source code for our simulator and market-making agents is available at \url{https://github.com/JJJerome/rl4mm}.

\section{Preliminaries}


\subsection{Experimental setup}\label{sec:experimental-setup}

Our empirical evaluation uses 50 levels of limit order book data provided by
LOBSTER\footnote{\url{https://lobsterdata.com/}}. The data is replayed in a
custom-built data-driven high-fidelity simulator that integrates the historical
orders with orders placed by test agents.
We used data from the first two weeks of March 2022 for the following symbols
for our evaluation. 


\begin{table}[!ht]
	\setlength\belowcaptionskip{-10pt}
    \centering
	\resizebox{\linewidth}{!}{
    \begin{tabular}{llll}
        \toprule\toprule
		Ticker & Description & Exchange & Sector\\
        \midrule
		\ticker{AXP}  & American Express & NYSE & Finance \\
		\ticker{BA}   & Boeing Company & NYSE & Industrials \\
		\ticker{BAC}  & Bank of America Corp.\ & NYSE & Finance \\
		\ticker{CAT}  & Caterpillar, Inc.\ & NYSE & Industrials \\
		\ticker{GE}   & General Electric & NYSE & Consumer Discretionary \\
		\ticker{HPQ}  & HP Inc.\ & NYSE & Technology \\
		\ticker{IBM}  & Int.\ Business Machines Corp.\ & NYSE & Technology \\
		\ticker{JNJ}  & Johnson \& Johnson & NYSE & Health Care \\
		\ticker{JPM}  & JP Morgan Chase \& Co.\ & NYSE & Finance \\
		\ticker{KO}   & Coca-Cola Company & NYSE & Consumer Staples \\
		\ticker{MMM}  & 3M Company & NYSE & Industrials \\
		\ticker{TXN}  & Texas Instruments Inc.\ & NASDAQ-GS & Technology \\
		\ticker{WMT}  & Walmart Inc.\ & NYSE & Consumer Discretionary \\
        \bottomrule\bottomrule 
    \end{tabular}
	}
    \bigskip
\caption{Tickers for our empirical evaluation.}
    \label{tab:tickers}
\end{table}

\vspace*{-0.8cm} 

\subsection{The market-replay gym environment}\label{sec:gym_env}
To test our agents, we created a gym environment which mimics an exchange with
price-time priority. The environment allows the agent to interact periodically
and place limit orders, market orders, cancellations or deletions and accrue
cash and inventory of the traded asset. Between these interaction times, ultra
fine-grained (Nasdaq or NYSE) order data is replayed and the state of the order
book is updated. By allowing the agent to place their own orders and interact
with the order book directly, certain properties of real order books naturally
arise. For example, the agent faces market impact when placing market orders and
``walking the book''.

The mechanism of running an episode is as follows:
\begin{enumerate}[leftmargin=0.5cm]
	\item Choose a random start time 
		for the episode such that the entire episode occurs within the random
		trading day.

	\item Initialise the order book using the top (50 levels) of the historical
		order book at that point in time.

	\item Let the agent choose their desired order distribution (levels and
		volumes) in the order book.

	\item Using the desired and existing orders, create orders that will achieve
		the desired orders. This may require cancelling orders from levels 
		with too much volume, or placing limit orders. In addition, 
		the agent may wish to place a market order to clear inventory
		if it is outside of some threshold amount.

	\item Replay the historical orders that occurred between the current time step and the next time step. In doing so update the agent's cash and inventory of the traded asset and track profit and loss (PnL) and any other desired quantities.
	\item Repeat steps (3)-(5) until the episode is finished.
\end{enumerate}

The strengths and weaknesses of using market replay 
are discussed in \cite{Balch19}. 
Most notably, the main weakness of using market replay, compared to an
interactive method such as an agent-based model, is the lack of adaptiveness of
the future order flow to orders placed by the agent. In an 
agent-based simulator it is possible to include (exogenous) agents that
react to the state of the order book when modified by the internal agent.
However, agent-based models are notoriously hard to calibrate and so, whilst
they possess reactivity to agents' actions, their realism is questionable.

In contrast, market-replay simulators are highly realistic. If the agent chooses
not to place any orders in the book, then the evolution of the order book in the
gym environment is, by definition, perfect. Furthermore, if we are willing to
make the assumption that an agent placing small orders does not affect the
future order flow too dramatically, then market replay should also accurately
model the case with agent interaction. This is important to keep in mind when
allowing the agent to interact with the book -- if they constitute too large a
proportion of the order flow volume then such an assumption is clearly violated.

\subsection{Beta volume profiles}
\label{subsec:beta}

A market maker will continually place a set of bids and a set of asks.
We first set a parameter, \totalvolume, which specifies the total amount of volume
that the market maker will place on each side of the market. 
Having this quantity fixed is not as restrictive as it sounds, since the market
maker can skew the volume distribution across prices so that an arbitrary
proportion of volume can be far from the current best prices and thus very
unlikely to be executed.

Let $f(x) = \frac{x^{\alpha-1}(1-x)^{\beta-1}}{\textrm{B}(\alpha, \beta)}$ be the probability density function of a beta distribution with parameters $\alpha$ and $\beta$, where $\textrm{B}(\cdot, \cdot)$ is the beta function. It can then easily be confirmed by a change of variables that the \textit{scaled beta distribution} with probability density function 
\begin{eqnarray}\label{eq:scaled_beta_pdf}
	g(x) = \frac{1}{{\nlevels}}f\left(\frac{x}{\nlevels}\right)
\end{eqnarray}	
and support $[0,\nlevels]$ is also a probability distribution.

We may therefore represent the distribution of the two sets of orders (for bids
and asks) with two scaled beta distributions, 
\begin{align*}
\bidorders\sim\scaledbetad(\nlevels,\abid,\bask),\\
\askdorders\sim\scaledbetad(\nlevels,\aask,\bask),
\end{align*}
where \nlevels is a fixed integer than specifies the support of the scaled 
distribution, which will correspond to the set of (contiguous) price levels at
which the agent will quote.\footnote{All prices are relative prices measured
away from the touch. If $p^*_b$ is the best bid, $p^*_a$ is the best ask and
$\Delta$ is the minimum tick size, then the bid prices that the agent quotes at
are $p^*_b - (i-1)\Delta$ and the ask prices are $p^*_a + (i-1)\Delta$ for
$i\in\{1,2,\dots,\nlevels\}$.}

We will see in Section \ref{sec:beta-generalisation} that this approach generalises the three natural approaches taken in
the literature -- the single order policy, the ladder strategy and market making
at the touch --  whilst permitting significantly more flexibility.\footnote{To
	fully generalise the ladder strategy significantly, an extra parameter
	$\minquote$ needs to be defined as the lowest level at which the agent places orders. We remark in Section \ref{sec:fixed-beta} that we may set $\minquote=0$ without losing much generality.}

\subsection{Implementing a beta volume profile}

To implement a volume profile in practice, two steps must be taken. First, the
\textit{continuous} scaled beta distribution must be \textit{quantised} to
represent orders of integer size at integer levels. Next, the quantised
distribution must be converted into orders and sent to the exchange.

\paragraph{Quantising the beta distribution.} To decide how much volume to place
at each level, the agent evaluates $g$ from \eqref{eq:scaled_beta_pdf} at $x_i =
i-0.5$ for $i\in \mathcal{I}\coloneqq\{1,2,\dots, \nlevels\}$ and normalises the results so that they sum to $1$.
This approximates the continuous \scaledbetad~distribution by a discrete
distribution with support $\mathcal{I}$. Then, the agent scales this
result by their \totalvolume and rounds the volume placed at each level to the
nearest integer.\footnote{This might cause the total volume of orders that the
agent places in the book to deviate slightly from the parameter \totalvolume,
however it is equal on average.} This is the agent's desired \textit{limit order
profile}.

\paragraph{Converting the quantised distribution into orders}
The simulator keeps track of the agent's `internal order book' -- the limit
orders belonging to the agent that are currently active. At each time step, the
agent converts this order book into two vectors representing the current limit
order profile held on the bid and ask sides of the order book. The agent then
subtracts these vectors from the desired limit profile to get a signed volume
difference which is converted into orders. If the difference is positive, then
the agent places a limit order for that amount. If the order is negative, then
the agent places cancellations for limit orders in the queue at that
level. Here, we assume that volume is cancelled from the back of the
queue first. This choice is based upon the assumption that, on average, having
orders nearer the front of the queue is beneficial to the agent.

\begin{figure*}
	\centering
	\begin{subfigure}[b]{0.24\linewidth}
		\centering
		\includegraphics[width=\linewidth]{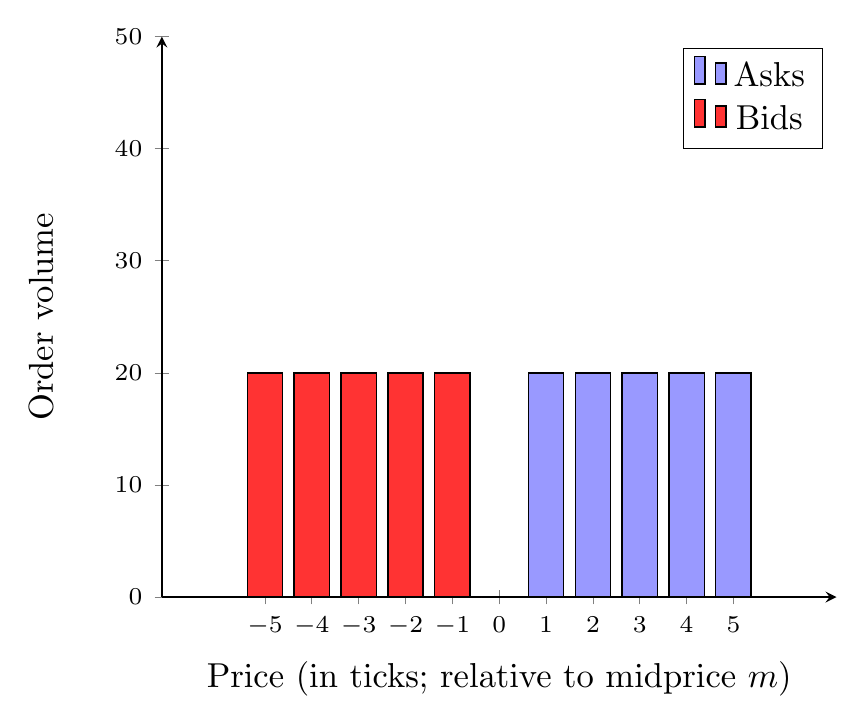}
		\caption[]%
		{{\tiny \quad$(\abid, \bbid) = (1,1)$,\ \ $(\aask, \bask) = (1,1)$}}    
	\end{subfigure}
	\hfill
	\begin{subfigure}[b]{0.24\linewidth}  
		\centering 
		\includegraphics[width=\linewidth]{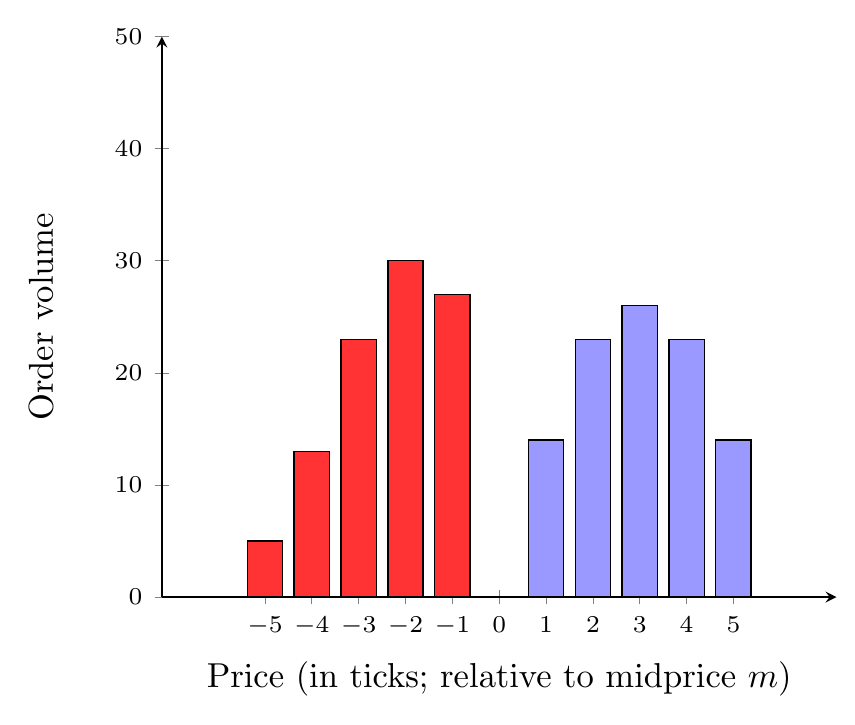}
		\caption[]%
		{{\tiny \quad$(\abid, \bbid) = (2,5)$,\ \ $(\aask, \bask) = (2,2)$}}   
	\end{subfigure}
        \hfill
	\begin{subfigure}[b]{0.24\linewidth}   
		\centering 
		\includegraphics[width=\linewidth]{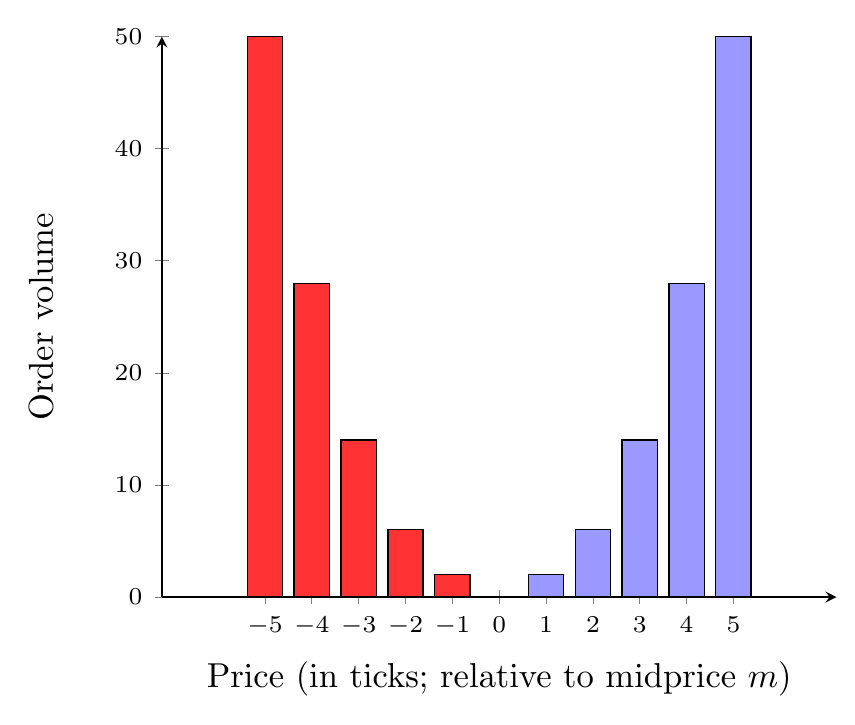}
		\caption[]%
		{{\tiny \quad$(\abid, \bbid) = (5,1)$,\ \ $(\aask, \bask) = (5,1)$}} 
	\end{subfigure}
	\hfill
	\begin{subfigure}[b]{0.24\linewidth}   
		\centering 
		\includegraphics[width=\linewidth]{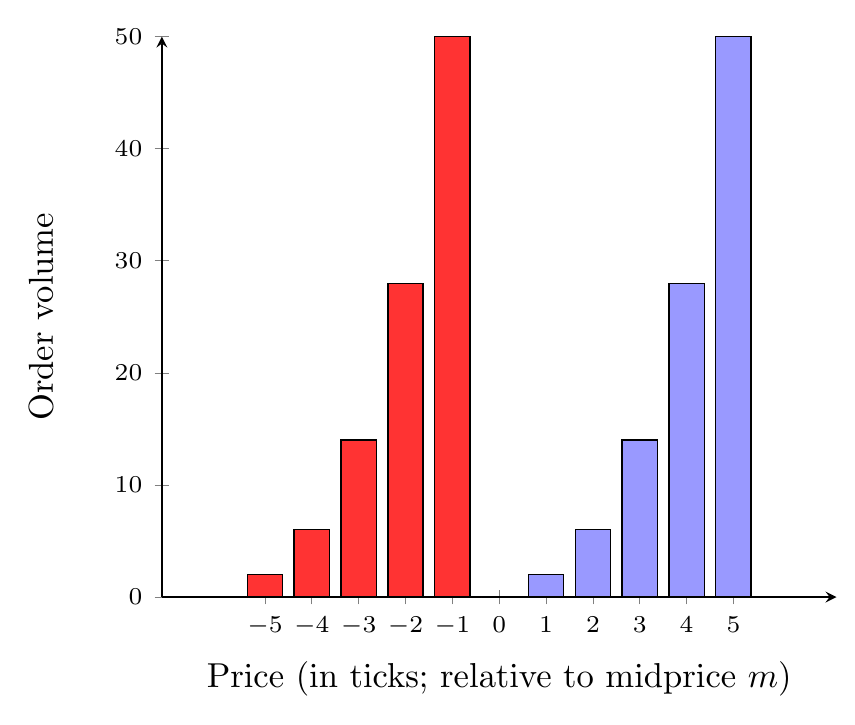}
		\caption[]%
		{{\tiny \quad$(\abid, \bbid) = (1,5)$,\ \ $(\aask, \bask) = (5,1)$}}  
	\end{subfigure}
	\caption[Examples of action choices]
	{\small A range of different action choices: a) is a ladder strategy; b) is the example 
	from Figure~\ref{fig:lob}; c) skews the volume profile on both sides away from the best 
prices; d) skews bids towards the best prices and asks away from the best prices, so would 
be a sensible action to take to try and redress a negative inventory.}
	\label{fig:examples}
\end{figure*}

\subsection{Representing a beta distribution by its mode and concentration.} The
use of $\alpha$ and $\beta$ to specify a beta distribution are arguably not as
natural as using other distributional statistics such as the mean, variance, or
mode, which are more interpretable. In this paper, for our inventory-based policy
we use the \emph{mode} and \emph{concentration} to specify beta distributions
that correspond to the agent's desire for a mean-reverting inventory.
\footnote{It is also possible to parameterise the distribution by its mean and
variance. For a chosen mean $\mu$ and variance $\sigma^2$, $\alpha$ and $\beta$
are given by $\alpha = ((1-\mu)\sigma^{-2} - \mu^{-1})\mu^2$ and
$\beta=\alpha(\mu^{-1} - 1)$. However, when choosing these parameters, it is
necessary that $\sigma^2 <\mu(1-\mu)$. This is problematic for an agent that
wishes to \textit{learn} these parameters, since the shape of the control space
is unusual.} 
The
following equations relate the mode and concentration to $\alpha$ and $\beta$.
\begin{definition}[Mode and concentration]
	The concentration, $\kappa$ of a beta distribution is defined as
	$\kappa = \alpha + \beta.$
	When $\alpha, \beta > 1$, which we will assume throughout this paper,
	the mode, $\omega$ is $\omega = \frac{\alpha-1}{\alpha + \beta - 2}$.
	Given suitable $\omega$ and $\kappa$, the corresponding $\alpha$ and $\beta$
	are given by:
	\begin{equation}
	\label{eq:convert-omega-kappa}
	\alpha =\ \omega (\kappa - 2) + 1, \beta  =\ (1 - \omega) (\kappa - 2) + 1.
	\end{equation}
	
	In a slight abuse of notation, we use the parameter $\omega$ of the underlying $\betad$ distribution to parametrise $\scaledbetad$. This can be thought of as the \textit{proportion} of the $\nlevels$ at which the agent wants their mode, so that the mode for $\scaledbetad$ is $\omega\cdot\nlevels$.
\end{definition}

\section{Beta policies for market making} \label{sec:beta_policies_for_mm}

We assume that $\nlevels$ is fixed during an episode. Then, at each time step,
the basic beta action is defined by the following 4-tuple, which defines two
scaled beta distributions: 

\begin{equation}
\label{eq:fourtuple}
a=(\alpha^\bid, \beta^\bid, \alpha^\ask, \beta^\ask)
\end{equation}

We also consider restrictions, where the concentration of the beta distribution
is set as a constant and the mode is then varied, with
\eqref{eq:convert-omega-kappa} used to recover the corresponding $\alpha$ and
$\beta$. This reduces the action space of the agent to be 2 dimensional and
resembles a smoothed version of the \textit{single price-level} action space.

\subsection{Special cases of scaled beta distributions}\label{sec:beta-generalisation}

Before discussing one of the main advantages of beta policies -- the 
ability to alter one's skew while maintaining queue position -- we explain
how beta policies generalise the three types of prominent action spaces from
the literature. 
As mentioned above, to emulate a single price-level policy
(Section~\ref{sec:single-price-level}), one can set the variance of the beta
distribution close to zero. 
To recreate a ladder strategy one uses \betad(1,1).
Finally, to emulate market making at the touch, to place volume at the 
touch the agent would 
choose parameters $\alpha =1, \beta \gg 1$ so
that the scaled beta distribution reduces to a Dirac delta at the first level.
Similarly, if the agent wants to not place any volume at the touch, they can
choose $\beta=1, \alpha\gg1$. 
Provided that $\nlevels$ is large enough, this amounts to placing orders far
away from the touch. 
In particular, they are highly unlikely to ever get executed on that side.

\subsection{Maintaining queue position}
\begin{figure}[h]
	\centering
	\includegraphics[width=0.78\linewidth]{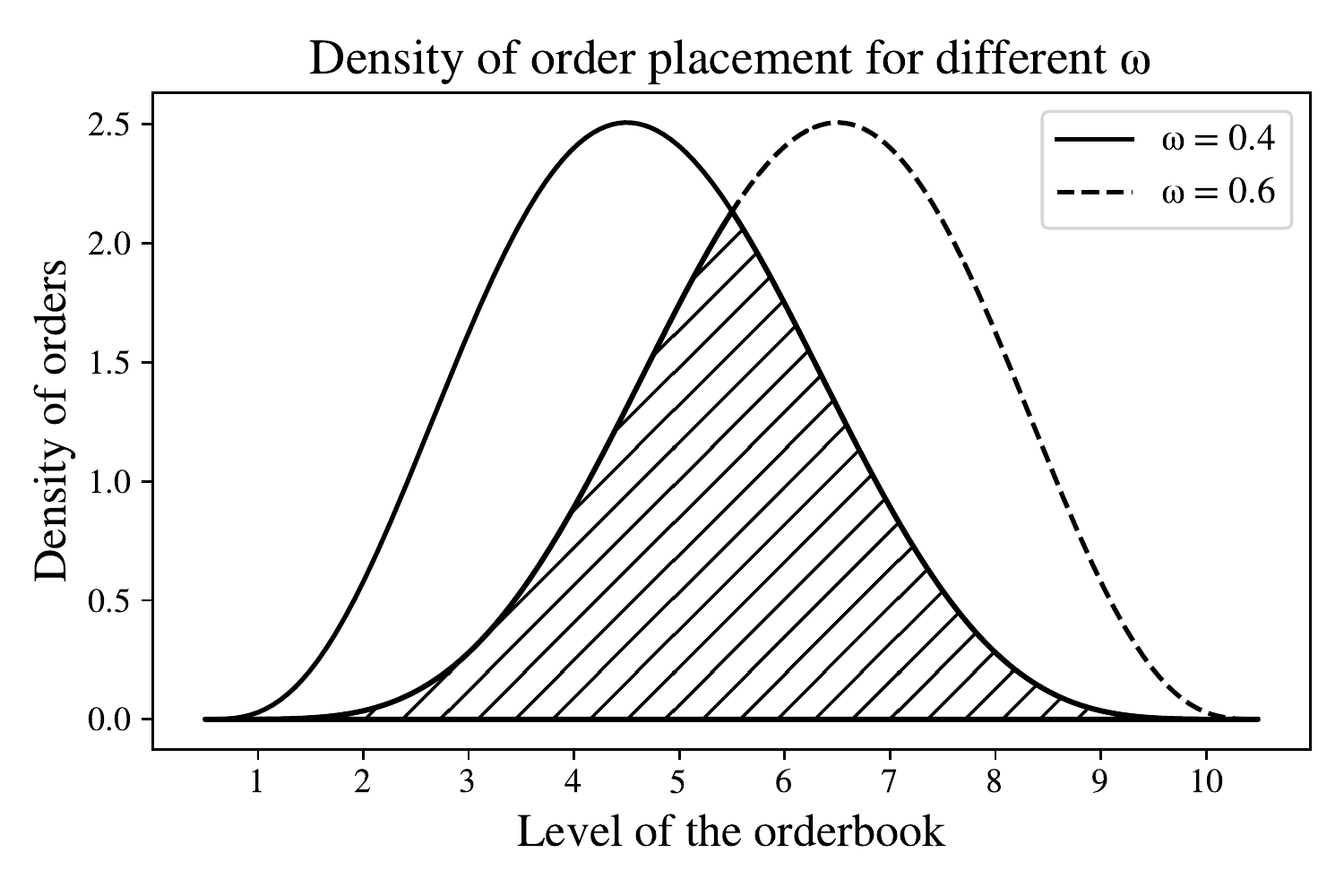}
	\caption[Book density for two different values of the mode $\omega$ with the concentration $\kappa=10$ fixed.]
	{\small Book density for different values of the mode $\boldsymbol\omega$, with the concentration $\boldsymbol{\kappa=10}$ fixed.}
	\label{fig:bookdensity}
\end{figure}

As described in Section \ref{sec:single-price-level}, when placing orders at a
single level, the agent must frequently cancel and replace orders at the levels
they have chosen in that time step. Due to the price-time priority mechanism
implemented by most major exchanges, this causes them to lose their queue
position and join the back of the queue at the new level. However, this
phenomenon is much weaker for scaled beta policies. This is because much of the
agent's volume is unaltered when adjusting the desired mode of the distribution
and so they may continuously update their action tuple, whilst not constantly
requiring that they leave and rejoin the queue. This is illustrated in Figure
\ref{fig:bookdensity}, representing the change in order volume for an agent that
wanted to change $\omega = 0.4$ to
$\omega = 0.6$ in their $\scaledbetad$ order distribution, whilst maintaining a concentration of $\kappa = 10$. In
particular, the volume in the shaded region is unaltered so that they do not
lose their queue position on it.

\subsection{Why the beta distribution?}

In this section, we explain why we chose the scaled beta distribution in terms of 
its advantages compared to prominent alternatives.
For our application, its main benefits are that it:
\begin{enumerate}
	\item allows an independent choice of the mean and
		variance;\footnote{Technically, the mean and variance must be possible
			for the bounded random variable, and given a bounded support and
			choice of mean, the variance will be constrained. However, compared
			to alternative distributions the constraints are weak}
	\item only requires two intuitive parameters $\omega$ and $\kappa$;
	\item has compact support over a fixed number of levels.
\end{enumerate}

For (1), the flexibility in terms of choosing the mean and variance
independently allow us to generate a wide variety of volume profile ``shapes''
with the scaled beta distribution; in particular we can emulate both ladder
strategies (high variance) and single-price level actions (low variance). In
contrast, alternatives such as the binomial distribution or the beta-binomial
distribution, have strong constraints on the variance.
The binomial distribution has a fixed variance for a given mean. 
The beta-binomial relaxes this slightly, but still constrains the variance within a
fixed interval for a given mean. 
This would mean that certain low variance profiles that an agent may want to use
would be unavailable to them.


For (2), we remark that having just two intuitive parameters, not only aids
interpretability, but also simplifies any learning problems for which the volume
profile parameters are the output. An alternative would be to use a distribution
given by a polynomial or a neural network, but this would require significantly
more parameters that are less interpretable.\footnote{In
\citet{ganesh2019reinforcement}, the authors use a fitted quadratic function to
represent the exchange order book shape. However their goal here is not to
represent an agent's policy with this approach, but rather to simulate the
exchange order book from which they take liquidity to hedge a position gained
from trading in over-the-counter markets.}

For (3), we remark that for computational and practical reasons it would be 
undesirable to have a policy that could place limit orders at arbitrarily many 
different price levels. It would be possible to use a distribution with unbounded
support and then truncate; the scaled beta distribution's bounded 
support means that this is not necessary and we naturally place orders at no more 
than $\nlevels$ from the touch.
It is for this reason that we prefer the scaled beta distribution to 
Gaussian, Poisson, gamma, and lognormal distributions.


\subsection{A comparison of fixed beta policies}\label{sec:fixed-beta}


A ladder strategy is parametrised by its two end points. We consider intervals
of 10 units of the minimum tick size and vary the ``best price'' of the ladder.
Here, we vary the best price between -2 ticks, which means go within the 
spread by 2 ticks where possible, up to~7, which means start at 
7 ticks outside of the touch.

We tested a variety of fixed ladder strategies using 60 randomly drawn hour-long episodes on the 13 assets over the timeframe outlined in Section \ref{sec:experimental-setup}. To compute ``returns'' we divided the profit or loss by the 
first price of the respective asset during the time period, and we report the 
mean and standard deviation of these $60 \times 13$ ``returns'' in each row. We
also report the number of profitable tickers in terms of the total profit and
loss (so the maximum possible in this column is 13). 
Table~\ref{tab:ladder} shows that the basic ladder strategy never has a positive
mean return for any minimum quote level, with the largest number of profitable
tickers being 4. 

\begin{table}[h]
\centering
\resizebox{\linewidth}{!}{ 
	\begin{tabular}{r|r|r|r|r|r}%
$\alpha^{\rm bid}=\alpha^{\rm ask}$ & $\beta^{\rm bid}=\beta^{\rm ask}$ & min quote & \# profitable & mean return & std returns\\\hline
1 & 1 & $-1$ & 0 & $-$1.96  & 8.80\\ \hline
1 & 1 & $-2$ & 3 & $-$2.58  & 20.13\\ \hline
1 & 1 & 0 & 0 & $-$1.06  & 5.22\\ \hline
1 & 1 & 1 & 0 & $-$0.68  & 3.57\\	\hline
1 & 1 & 2 & 2 & $-$0.48  & 3.27\\ \hline
1 & 1 & 3 & 4 & $-$0.28  & 2.69\\ \hline
1 & 1 & 4 & 0 & $-$0.28  & 3.08\\ \hline
1 & 1 & 5 & 1 & $-$1.07  & 6.30\\ \hline
1 & 1 & 6 & 2 & $-$0.24  & 2.87\\ \hline
1 & 1 & 7 & 1 & $-$0.33  & 2.83\\
	\end{tabular}
} 
\caption{Performance of ladder strategies for different minimum quote levels, from two ticks inside the spread 
(when possible) to seven ticks from the touch.}
\label{tab:ladder}
\end{table}

We next allow non-uniform beta actions, and sweep over some simple parameter
combinations. In these experiments we always used minimum quote level 0. Recall
that a non-uniform beta action can place its mode at any of the possible
10 quote levels. This means that the generality gained from adding a different
minimum quote level is minor. We further ran some sweeps in which the minimum
quote level was allowed to vary along with the paramters $\alpha$ and $\beta$,
but found that the optimal value for the minimum quote level was close to zero.
We do not include these tables due to space constraints.
\looseness=-1

The results of the sweep over fixed parameter beta policies is given in
Table~\ref{tab:fixed}. Here, we see that the number of profitable tickers
increases and for two non-uniform beta policies a positive mean return is obtained. 
The profitable
fixed action policies were symmetric $\scaledbetad(10,1,2)$ and $\scaledbetad(10,1,5)$ policies,
which correspond to $\betad(1,5)$ and $\betad(2,5)$ respectively and therefore
are skewing the volume profile \emph{towards} the best prices.

While we have demonstrated the potential of non-uniform beta policies, it is
worth noting that their returns have large standard deviations (a simple measure of
risk). We will see in Section~\ref{sec:market-orders} that this is due to a lack
of inventory control demonstrated by such policies. We address this in the rest
of this section.

\begin{table}[h]
\centering
\resizebox{\linewidth}{!}{ 
	\begin{tabular}{r|r|r|r|r|r}%
$\alpha^{\rm bid}=\alpha^{\rm ask}$ & $\beta^{\rm bid}=\beta^{\rm ask}$ & min quote & \# profitable & mean return & std returns\\\hline
1 & 1 & 0 & 0 & $-$1.06 & 5.22\\ \hline
1 & 2 & 0 & 6 & 0.58 & 16.04\\ \hline
1 & 5 & 0 & 7 & 0.06 & 35.66\\ \hline
2 & 1 & 0 & 3 & $-$3.10 & 18.92\\ \hline
2 & 2 & 0 & 0  & $-$1.39 & 5.92\\ \hline
2 & 5 & 0 & 3 & $-$1.42 & 16.70\\ \hline
5 & 1 & 0 & 3 & $-$3.06 & 29.63\\ \hline
5 & 2 & 0 & 7 & $-$0.30 & 14.67\\ \hline
5 & 5 & 0 & 0  & $-$1.40 & 6.76 
	\end{tabular}
} 
\caption{Performance of fixed actions, including ladder strategies, for minimum quote level 0 (at the touch).}
\label{tab:fixed}
\end{table}

\vspace*{-0.5cm}

\subsection{Controlling inventory with market orders}\label{sec:market-orders}

\begin{figure}[h]
	\centering\hspace{-12pt}
	\begin{subfigure}[b]{0.795\linewidth}
		\centering
		\includegraphics[width=\linewidth]{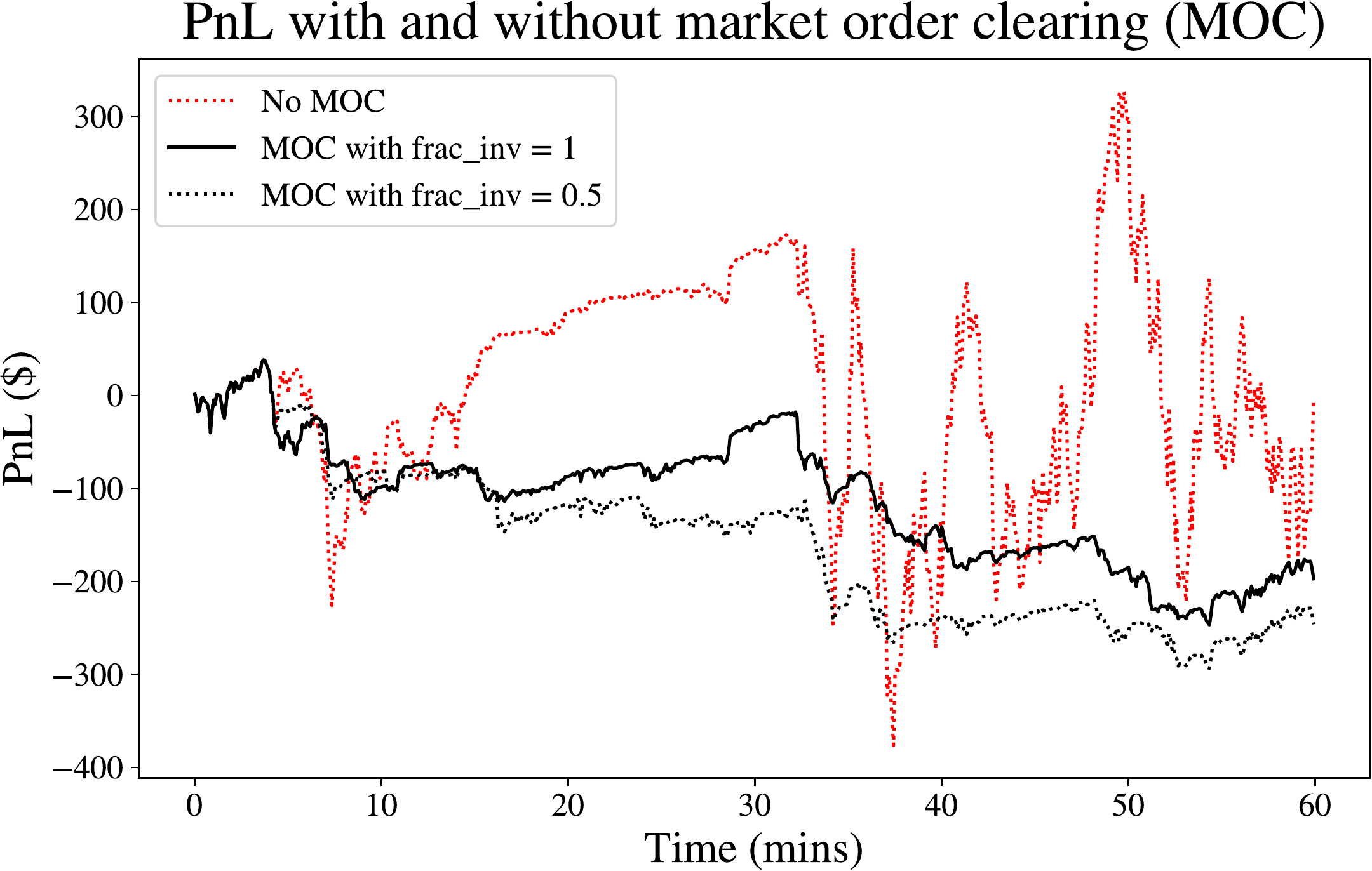}
	\end{subfigure}
	\begin{subfigure}[b]{0.85\linewidth}  
		\centering
		\vspace{10pt}
		\includegraphics[width=\linewidth]{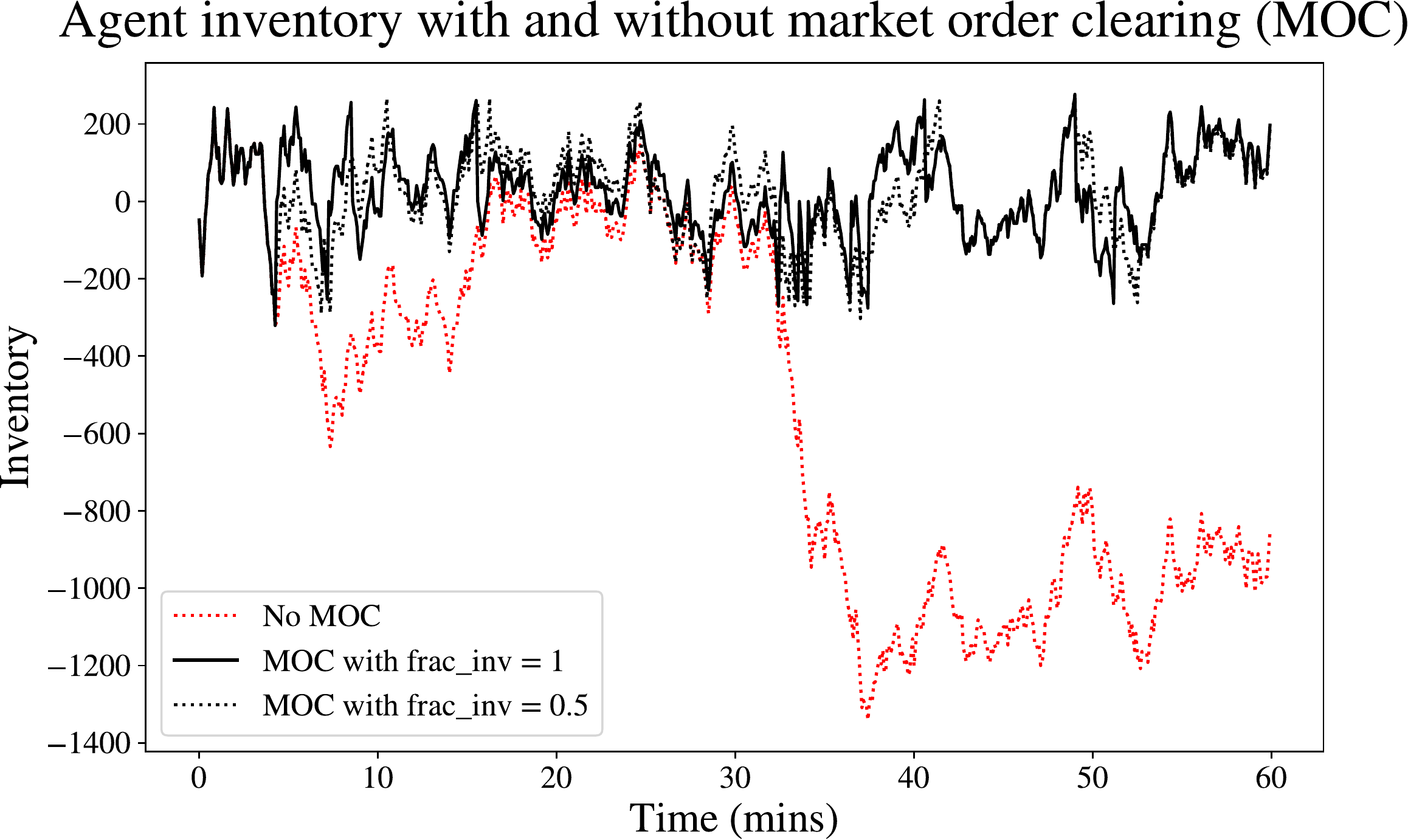}
	\end{subfigure}
	\caption[]{\small A comparison of the PnL and the inventory for an agent with
		the optimal fixed action $\boldmath{\abid = \aask = 1}$ and $\boldmath{\bbid
			= \bask = 2}$ from the sweep in Section~\ref{sec:fixed-beta} and a variety of
		different market order inventory clearing policies. These plots are for 
		the ticker JPM on the first date of the data (2nd March 2022) between 10:30
		am and 11:30 am with $\boldmath{\maxinv=250}$.}
	\label{fig:market-vs-fixed}
\end{figure}

The main source of risk for a market maker comes from holding inventory during
adverse price swings. Therefore one of the main goals of a market maker is to
make sure their inventory remains within some reasonable bounds.
Figure~\ref{fig:market-vs-fixed} shows that the optimal fixed beta strategy
found in Section~\ref{sec:fixed-beta} accumulates a huge negative inventory,
causing it to face wild price swings. One of the options available to a market
maker is to place a market order to liquidate some of their position in the
asset.

This extra action of placing a market order (with size proportional to the
agent's inventory) was considered in \cite{spooner2018market}. In their
experiments they set the proportion to be 1, which means that the agent
liquidates their entire inventory when it becomes too large.
\looseness=-1

To add this option to our agent with a scaled beta policy controlled by
\eqref{eq:fourtuple}, we add two extra parameters: the first is the maximum
absolute inventory ($\maxinv$) that the agent is willing to hold, and if they
surpass this level they place a market order to reduce their absolute inventory;
the second is the fraction of their inventory that they liquidate in such a
situation. Since our agents have a continuous action space, we decided that it
was natural to turn the impulse control problem of choosing when to liquidate
into a continuous control problem of choosing an appropriate risk limit defined
in terms of the absolute inventory. Therefore, the agent's action 4-tuple in \eqref{eq:fourtuple} becomes a 6-tuple,
\begin{equation}
\label{eq:sixtuple}
a = (\alpha^\bid, \beta^\bid, \alpha^\ask, \beta^\ask, \maxinv, \invfrac).
\end{equation}

Figure~\ref{fig:market-vs-fixed} compares three different market order clearing policies (no market clearing, market clearing with $\maxinv =1000$ and $\invfrac =1$, and market clearing with $\maxinv=1000$ and $\invfrac=0.5$). In particular, the strategy with $\invfrac=1$ manages inventory very well but has worse returns than that with $\invfrac=0.5$. Both greatly outperform the agent with no market clearing. However, the agent still pays a cost for this risk management in the form of crossing the spread. In the next section, an alternative strategy is introduced, which manages to control inventory, whilst only placing limit orders.

\subsection{An inventory-driven policy}\label{sec:inventory-driven-policy}

In the mathematical finance literature, it is well established that an inventory-aware 
strategy will skew its bid and ask according to its current inventory
level. In particular, an optimal agent trading in a financial market with
midprice process given by arithmetic Brownian motion and Poisson market order
arrivals (see \cite{avellaneda2008high, cartea2015algorithmic}) should skew the
midprice of their bid and ask quotes in the opposite direction to their asset
holdings; for example, if they hold a negative inventory, they should skew their
quotes so that their quoted midprice is higher than that of the market. Whilst
the models of \citet{avellaneda2008high} and \citet{cartea2015algorithmic} are
not necessarily very realistic, this intuitively makes sense since skewing in
this way makes it more likely that they will get filled on the side that brings
their inventory closer to zero, helping them to complete round-trip trades and
encouraging mean-reversion of their inventory (to zero).

It is this strand of literature that inspires our inventory-driven beta policy.
In particular, the following parametric form ensures that the agent skews their
active orders in such a way as to encourage mean-reversion. One can see that the
parametric form given below implies that if the agent's inventory is close to
their maximum desired inventory, then $\omega^\bid$ is close to one (the mode of
their bid orders is far away to the midprice), and $\omega^\ask$ is close to
zero (the mode of their ask orders is close to the midprice). When inventory is
zero the distribution of the agent's orders will have a mode that is some
proportion $\defomega$ (of the \nlevels they quote at) into the book.

\noindent \includegraphics[width=\linewidth]{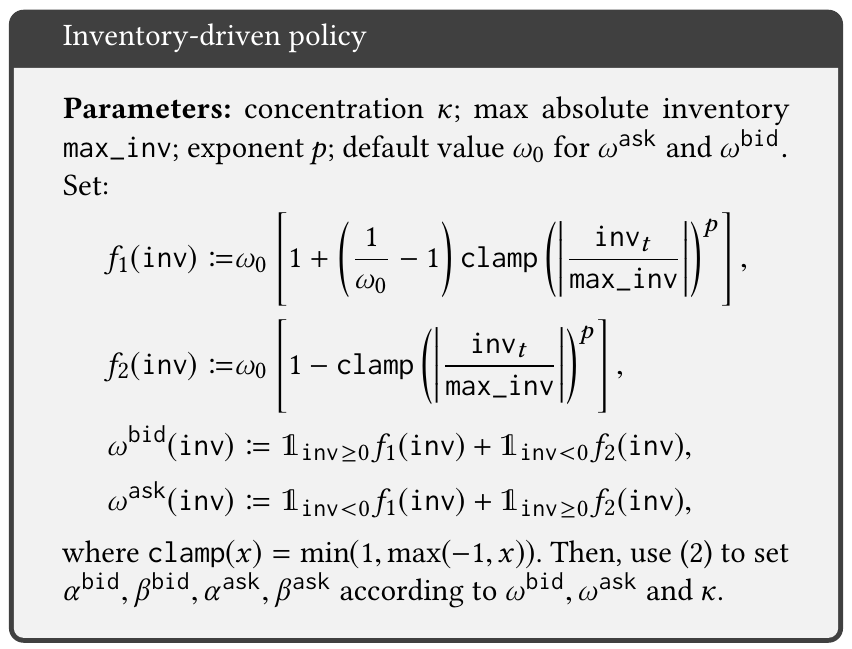}

%
%
%
%
%
%
%
%

\begin{figure}
	\centering
	\begin{subfigure}[b]{0.49\linewidth}
		\centering
		\includegraphics[width=\linewidth]{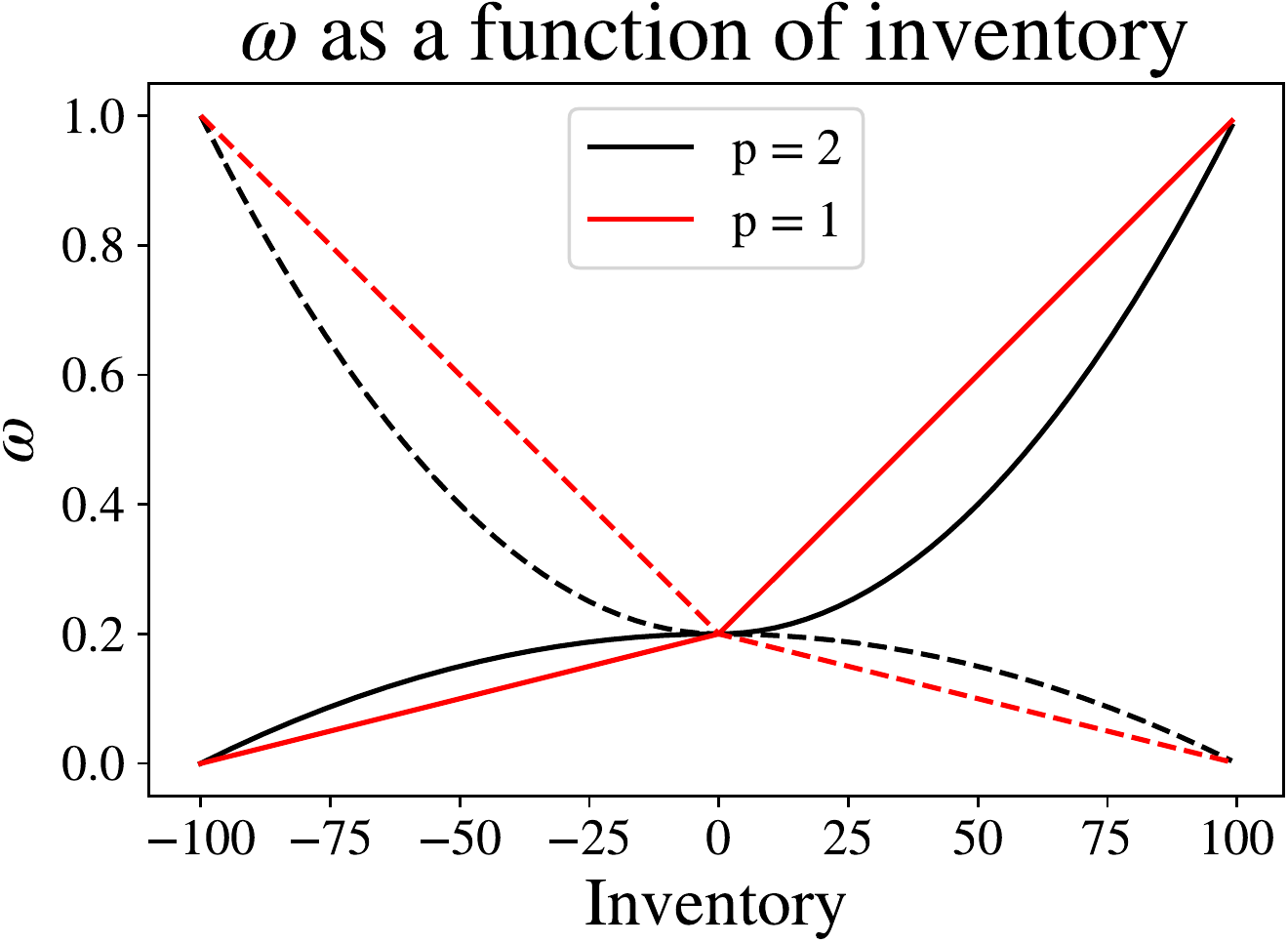}
	\end{subfigure}
	\hfill
	\begin{subfigure}[b]{0.49\linewidth}  
		\centering 
		\includegraphics[width=\linewidth]{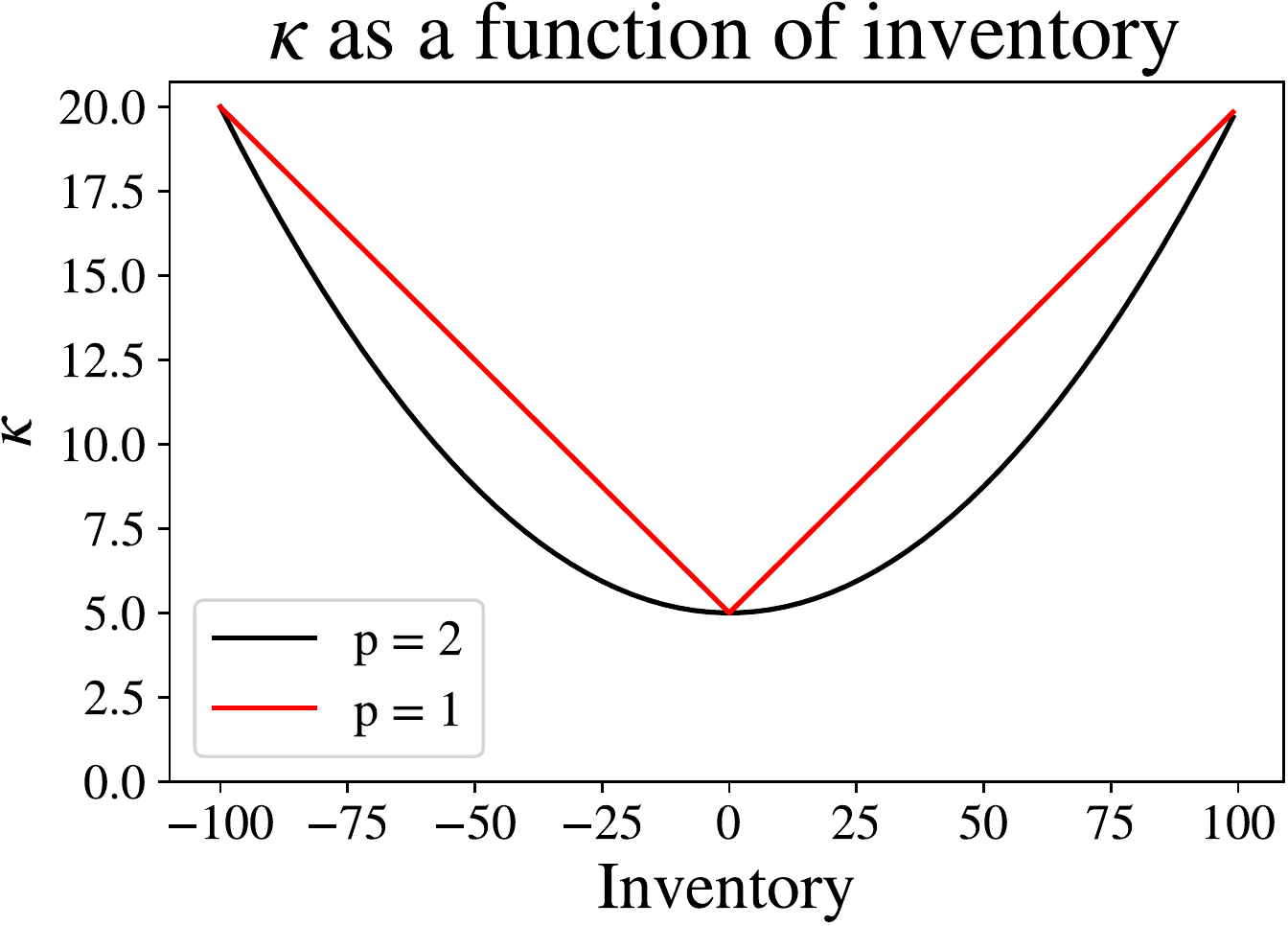}
	\end{subfigure}
	\caption[$\omega$ and $\kappa$ as functions of Inventory]
	{\small $\boldsymbol\omega$ and $\boldsymbol\kappa$ as functions of Inventory. Here, the dotted line in the left hand figure is $\boldmath\omega^\ask$ and the solid line is $\boldmath\omega^\bid$.}
	\label{fig:omegakappa}
\end{figure}

Here, the exponent $p$ controls the convexity of $\omega^{\ask}$ as a function
of inventory and generally takes a value $p\geq1$. Such superlinear dependence
of inventory on midprice skew is observed in the mathematical finance
literature~\cite[Chapter 10]{cartea2015algorithmic}. We further generalise to
the case when there exists a default value $\defkappa$ and a maximum value
$\maxkappa$ for $\kappa$. In this case, we define 
$$\kappa(\inv) \coloneqq (\maxkappa - \defkappa) \clamp\left(\left|\frac{\inv_t}{\maxinv}\right|\right)^p +
\defkappa.$$ 
The functions $\omega(\inv)$ and $\kappa(\inv)$ are plotted in
Figure \ref{fig:omegakappa} for $\defomega=0.2$, $\defkappa=5$, $\maxkappa=20$
and $p=2$.

\subsubsection{Mean reversion of inventory}
\begin{figure}
	\centering\hspace{3pt}
	\begin{subfigure}[b]{0.877\linewidth}
		\centering
		\includegraphics[width=\linewidth]{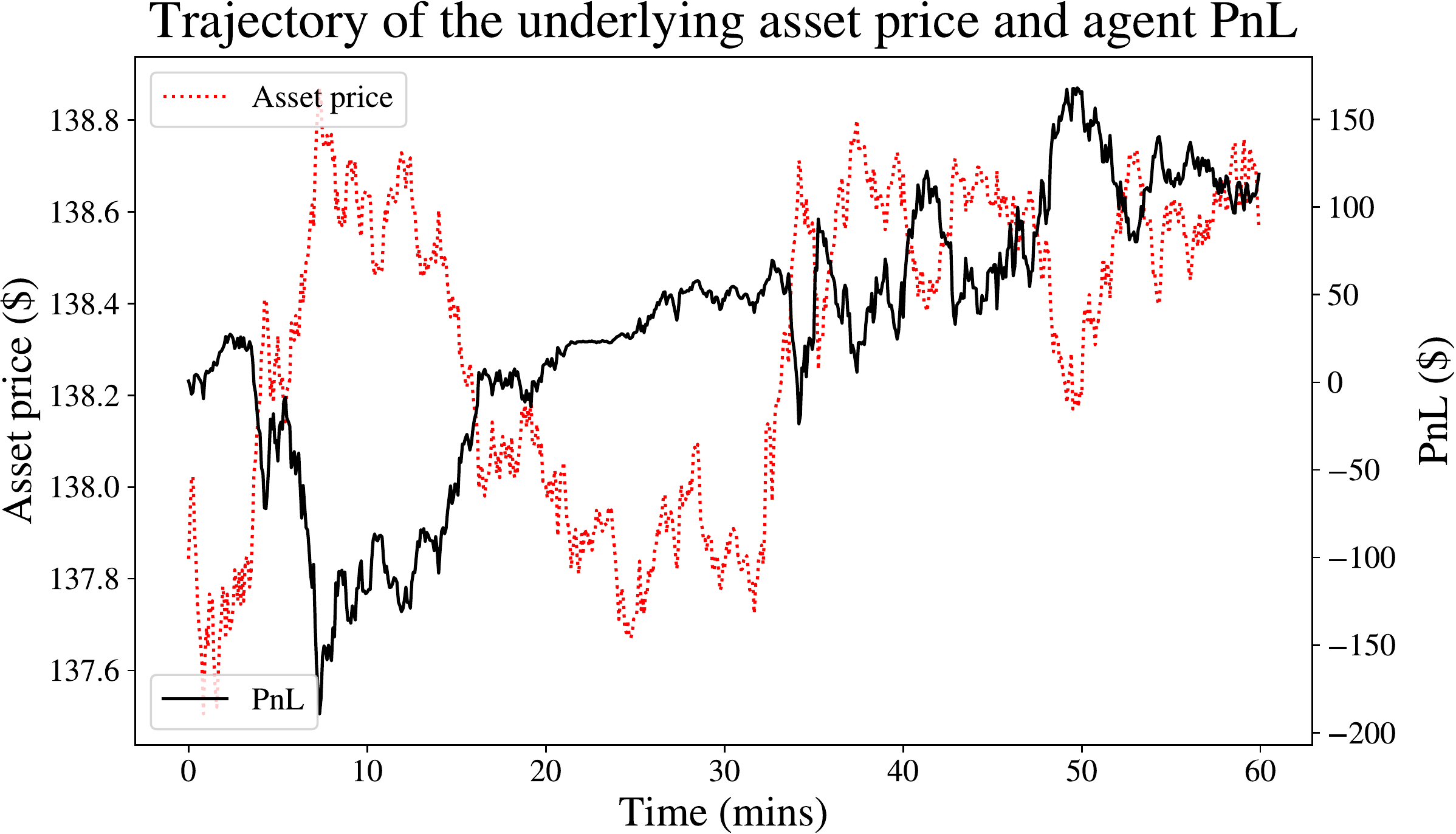}
	\end{subfigure}
	\begin{subfigure}[b]{0.85\linewidth}  
		\centering 
		\vspace{10pt}
		\includegraphics[width=\linewidth]{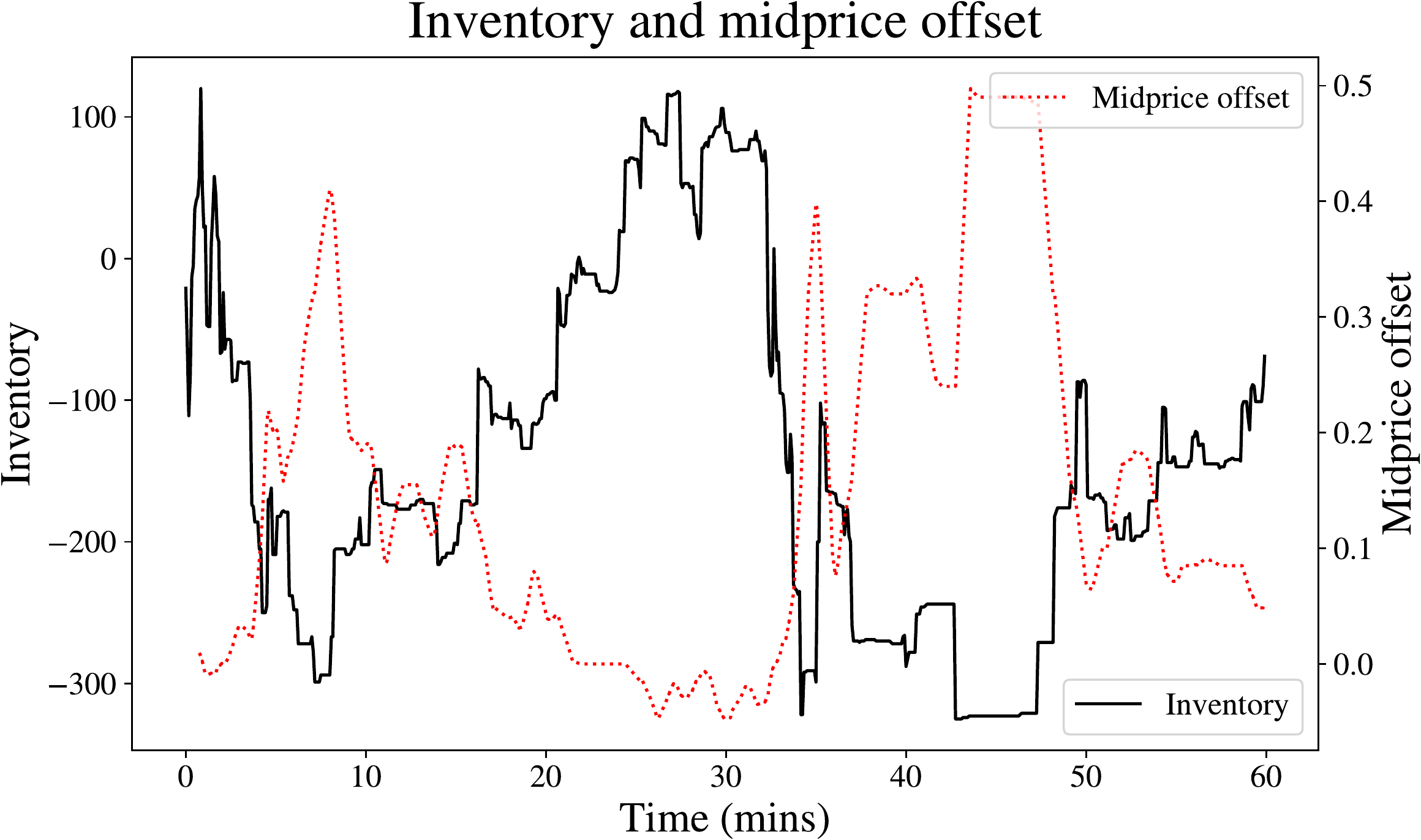}
	\end{subfigure}
	\caption{\small PnL for the inventory-driven agent compared with: the
		underlying asset price; the agent's inventory; and their midprice offset.
		These plots were generated for the ticker JPM and the same period as used in
		Figure \ref{fig:market-vs-fixed}.}
	\label{fig:inventory-midprice}
\end{figure}

The dependence of $\omega$ on inventory in the inventory-driven policy
encourages mean-reversion of the inventory by skewing the best bid and the
best offer. This can be seen in the bottom panel of Figure~\ref{fig:inventory-midprice}, in which the agent skews their midprice offset in
the opposite direction to their inventory. This
successfully induces a form of mean-reversion of the inventory levels.

\enlargethispage{-11pt}

Unlike the market order placing strategies of Section \ref{sec:market-orders},
the inventory-driven agent manages to do this without crossing the spread and
incurring a cost. Comparing the PnL graph of the inventory-driven
policy in the top panel of Figure~\ref{fig:inventory-midprice} with the PnL graph of 
the various market order agents in the top panel of Figure~\ref{fig:market-vs-fixed} 
(which are all calculated over the same period), we
see that such a strategy is effective. Finally, it is worth noting that the
agent manages to capture the spread and make a profit even though the market is
trending against them (they hold a negative inventory for the duration of the
episode, but the drift is positive).

\subsubsection{Tuned performance}
When running experiments, we found that such a strategy is not robust across
tickers and needed to be tuned. 
This could largely be due to the difference in relative tick size for
different assets -- an issue discussed in the subsequent section.
We tuned for the ticker JPM and, across 300 one-hour episodes, obtained the
following (in-sample) distribution for the ``returns''.

\medskip

{\begin{tabular}{cccc}
\begin{minipage}{0.35\linewidth}
\scalebox{0.98}{
\begin{tabular}{lr}
\textbf{mean} &    0.52\\ 
\textbf{std}  &    5.50\\ 
\textbf{min}  &  -38.47\\ 
\textbf{25\%} &     0.26\\ 
\textbf{50\%} &     1.20\\
\textbf{75\%} &     2.52\\
\textbf{max}  &   10.94\\
\end{tabular}
}
\end{minipage}%
&
\begin{minipage}{0.55\linewidth}
While it still suffered large losses on some episodes when it was
unable to control its inventory during highly trending periods, it
managed to make money on $78\%$ of episodes and had the best risk-reward profile
of all agents tested in the paper.
\end{minipage}%
\end{tabular}}

\subsection{Beta policies and tick sizes}

It is worth noting that the approach taken in this paper works best for assets
whose relative tick sizes lie `within a reasonable range'. If the relative tick
size is too large (the asset is \textit{tick-constrained}) then all activity
happens at the top level of the book.\footnote{There is also a time
component. If the agent acts every minute instead of every second, then 
by their perception of time, movements in the best bid and ask
occur more frequently. In this case, the approach of this paper is appropriate
again.} This means that, whilst there is benefit to maintaining queue position,
it is unnecessary to have the control over the volume profile afforded by the
beta distribution. The agent can just place block orders at a reasonable
quantity of levels as in a ladder strategy and then cancel volume at the
touch when they want to reduce fills.

If the relative tick size is too small, then the beta volume profile also needs
to be modified. This is because it is not sensible to post limit orders at every
level: firstly it is computationally expensive to track the agent's orders at so
many levels; secondly, if implemented in practice it would telegraph the agent's
presence to all of the other market participants.

A naive solution to the issues presented by small-tick stocks would be to only
post orders at uniformly-spaced price levels. However, this would not disguise
the agent's presence at all. A better solution would be to post orders at
randomly spaced price levels that are reasonably close to the desired uniformly
spaced price levels. This approach would require further development though,
since the random lattice would need to be updated more slowly than the agent
takes actions (or the agent would end up constantly placing and removing
orders).

\enlargethispage{-11pt}

The parameter \nlevels~is also sensitive to the relative tick size of the asset.
A more robust approach would be to use a beta distribution to characterise
volume profiles across price levels given in terms of basis points up to a value
\texttt{max\_bps}. After adding some randomisation to disguise the agent's
presence in the market and rounding to the nearest tick, this approach would be
robust across a range of asset sizes and could improve generalisability of
learnt strategies substantially. This is an easy addition to the code-base and
will be included in future work.

\section{Conclusions and further work}

We have introduced a new representation for market maker policies in limit order
book markets that derive limit order volume profiles from scaled beta
distributions.
This -- in contrast to most work in the AI4MM literature -- is a \emph{continuous
action space}, which makes it highly expressive.
It further encompasses the key special cases, of single orders, ladder strategies,
and market making at the touch, that have previously been studied in the AI4MM literature.
However, the approach is also significantly more general in terms
of the market maker's ability to simultaneously skew volume to favour orders on
one side of the market and maintain queue position, while using  
only a small number of easy-to-interpret parameters.


We believe that scaled beta volume profiles can form the foundation for 
sophisticated and performant market marking agents based on 
state-of-the-art learning approaches. For example, the following
directions for further work seem promising:

\begin{itemize}[leftmargin=0.35cm]
	\itemsep0mm
\item Bayesian optimization would be a natural approach to tune the 
	parameters of our inventory-based control policy since sampling trajectories is relatively costly.

\item \citet{AbernethyK13} treat different ladder strategies
	as ``experts'' and use multiplicative weights updates to dynamically
	select mixtures of these experts; it would be interesting to treat
	scaled beta policies in the same way. A key difference between the set of 
	ladder strategies and scaled beta volume profiles is that the former 
	set of strategies is finite whereas the latter is infinite; suitable 
	approaches for infinite mixtures exist, e.g.,~\cite{RasmussenG01}.
	In~\cite{AbernethyK13} they allow fractional volumes arising from mixtures
	of ladders; for realism, ideally a mixture approach would round
	volumes to integers, while still maintaining good performance. 

\item Arguably the most natural AI4MM approach to use with scaled beta volume
	profiles is \emph{Reinforcement Learning} (RL). With this approach, in each
	step scaled beta actions would be chosen based on features that describe
	both the market state (e.g. trendiness and volatility, and limit order book
	features that are known to have short term predictive value such as the book
	imbalance~\cite{bouchaud2018trades}) and the market marker's state (e.g.
	inventory). It would be extremely interesting to explore the use of
	state-of-the-art continuous control RL methods with a rich state space.

\end{itemize}


\begin{acks}
Jerome and Savani would like to acknowledge the support of a 2021 J.P.Morgan Chase
AI Research Faculty Research Award.
\end{acks}


\balance
\bibliographystyle{ACM-Reference-Format}
\bibliography{papers}

\end{document}